\documentclass[twocolumn,prX,aps,superscriptaddress,nofootinbib]{revtex4-1}
\usepackage[latin9]{inputenc}
\usepackage{mathtools}
\usepackage{amsbsy}
\usepackage{amstext}
\usepackage{appendix}
\usepackage{braket}
\usepackage{array}

\usepackage[unicode=true,pdfusetitle,
 bookmarks=false,
 breaklinks=false,pdfborder={0 0 1},backref=false,colorlinks=false]
 {hyperref}
 \usepackage{xcolor}
\usepackage{changepage}

\usepackage{perpage}
\MakePerPage{footnote}

\usepackage[english]{babel}

\hypersetup{
 bookmarksnumbered=false,bookmarksopen=false}
\makeatletter

\usepackage{amsmath}
\usepackage{graphicx}
\usepackage{amssymb}
\usepackage{txfonts,color}
\@ifundefined{textcolor}{}

{%
 \definecolor{BLACK}{gray}{0}
 \definecolor{WHITE}{gray}{1}
 \definecolor{RED}{rgb}{1,0,0}
 \definecolor{GREEN}{rgb}{0,1,0}
 \definecolor{BLUE}{rgb}{0,0,1}
 \definecolor{CYAN}{cmyk}{1,0,0,0}
 \definecolor{MAGENTA}{cmyk}{0,1,0,0}
 \definecolor{YELLOW}{cmyk}{0,0,1,0}
 \definecolor{ORANGE}{rgb}{1,0.5,0}
}

\makeatother

\begin{document}
\author{Pol Alsina-Bol\'{i}var}
\thanks{These authors have equally contributed to this work.}
\affiliation{Department of Physical Chemistry, University of the Basque Country UPV/EHU, Apartado 644, 48080 Bilbao, Spain}
\affiliation{EHU Quantum Center, University of the Basque Country UPV/EHU, Leioa, Spain}
\author{A. Biteri-Uribarren}
\thanks{These authors have equally contributed to this work.}
\affiliation{Department of Physical Chemistry, University of the Basque Country UPV/EHU, Apartado 644, 48080 Bilbao, Spain}
\affiliation{EHU Quantum Center, University of the Basque Country UPV/EHU, Leioa, Spain}
\author{C. Munuera-Javaloy}
\affiliation{Department of Physical Chemistry, University of the Basque Country UPV/EHU, Apartado 644, 48080 Bilbao, Spain}
\affiliation{EHU Quantum Center, University of the Basque Country UPV/EHU, Leioa, Spain}
\author{J. Casanova}
\affiliation{Department of Physical Chemistry, University of the Basque Country UPV/EHU, Apartado 644, 48080 Bilbao, Spain}
\affiliation{EHU Quantum Center, University of the Basque Country UPV/EHU, Leioa, Spain}
\affiliation{IKERBASQUE,  Basque  Foundation  for  Science, Plaza Euskadi 5, 48009 Bilbao,  Spain}
\affiliation{Corresponding author: jcasanovamar@gmail.com}

\title{J-coupling NMR Spectroscopy with Nitrogen Vacancy Centers at High Fields}

\begin{abstract}
A diamond-based sensor utilizing nitrogen-vacancy (NV) center ensembles permits the analysis of micron-sized samples through NMR techniques at room temperature. Current efforts are directed towards extending the operating range of NV centers into high magnetic fields, driven by the potential for larger nuclear spin polarization of the target sample and the presence of enhanced chemical shifts. Especially interesting is the access to J-couplings as they carry information of chemical connectivity inside molecules. In this work, we present a protocol to access J-couplings in both homonuclear and heteronuclear cases with NV centers at high magnetic fields. Our protocol leads to a clear spectrum exclusively containing J-coupling features with high resolution. This resolution is limited primarily by the decoherence of the target sample, which is mitigated by the noise filtering capacities of our method. 
\end{abstract}
\maketitle

\section{Introduction } 
Nuclear Magnetic Resonance (NMR) spectroscopy has established itself as a fundamental tool for exploring materials since its inception in the 1950s~\cite{Slichter78,levitt2013spin,tampieri2022a}.  This technique is essential for analyzing molecular structure and function. It finds extensive applications in the examination of macroscopic biological tissues and functional studies in medical imaging~\cite{,chary2008nmr,bottomley1984nmr,abragam1961principles}. However, the inherent low sensitivity of standard NMR spectroscopy restricts its application to bulky samples, typically requiring microliter volumes with conventional macroscopic coils or nanoliter volumes with microcoils~\cite{allert2022advances}. In this scenario, newly developed solid-state quantum sensors, open up new avenues for spectroscopic measurements with unprecedented sensitivity and spatial resolution, enabling studies at the cellular~\cite{neuling2023prospects} and  single-molecular level~\cite{sushkov2014all,shi2015single,lovchinsky2016nuclear}. In particular, quantum sensors consisting on NV ensembles have demonstrated their utility in conducting NMR spectroscopy on microscopic samples at room temperature, reaching sample volumes of picoliters~\cite{allert2022advances,glenn2018high,bucher2020hyperpolarisation,arunkumar2021micron}. The pursuit of extending the operating range of NV-based sensors into the high magnetic field regime is driven by the advantages offered by elevated fields. Namely, a larger thermal nuclear spin polarization and enhanced energy shifts that simplify data extraction.

The ability to access \textit{J-couplings }-- also known as \textit{scalar} or \textit{indirect couplings}-- is of particular interest in the actual spectroscopic framework. J-couplings are indirect intramolecular dipole-dipole interactions mediated through chemical bonds. They are responsible for complex splittings in NMR spectral lines, and the examination of J-coupling patterns allows scientists to obtain information regarding nuclear connectivity within a molecule. Hence, accessing J-couplings at the microscale regime would open new avenues in mass-limited scenarios for applications in, e.g., combinatorial chemistry, as well as in cell and membrane biology \cite{liu2017combinatorial}. Nevertheless, J-couplings are not trivial to detect since (i) They are independent of the externally applied field. Then, at high magnetic fields other field-dependent terms dominate the dynamics, and (ii) Long coherence times are required to resolve the spectral peaks corresponding to J-couplings owing to their weak value (typically, on the order of a few Hertz).

In this work, we present J-INSECT  (J-coupling Induced Nuclear Signal with Extended Coherence Time). This protocol enables the detection of homonuclear and heteronuclear J-couplings with NV ensembles (thus, accessing microscale samples) in the regime of high magnetic fields. J-INSECT integrates heterodyne techniques~\cite{schmitt2017submillihertz,munuera2023high,meinel2023high,boss2017quantm} and exerts noise filtering over the sample effectively enhancing its coherence.  This results in narrower peaks in the spectra, thereby increasing frequency resolution and thus enabling to resolve quantities on the order of a few hertz (such as J-couplings). Additionally, our method provides the capability to target specific types of couplings. In complex molecules, where numerous peaks may arise due to extensive couplings between nuclei, this allows for the selective targeting of only the couplings of interest, simplifying the spectra. We investigate the performance of J-INSECT in regimes including both a reduced and a large number of measurements, and analyse the impact of distinct noise sources (including the  presence of always-on J-couplings) demonstrating the feasibility of the method in common experimental conditions.

\begin{figure*}[]
\includegraphics[width=1 \linewidth]{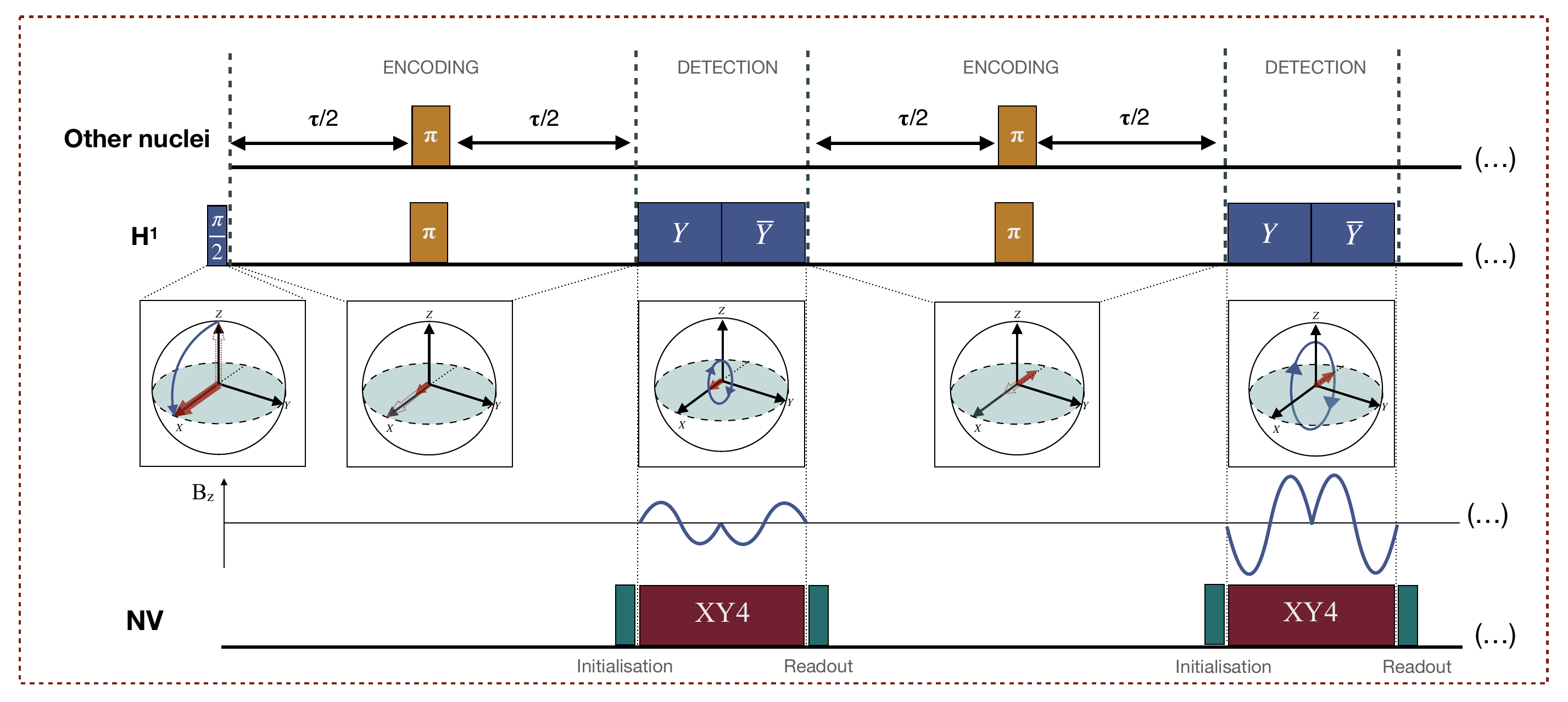}
\caption{{\bf Scheme of J-INSECT with RF and MW controls.} Each channel (black bold horizontal line) denotes the radiation pattern over each system constituent. The upper RF channel corresponds to the pulses over non-hydrogen nuclear species. In this respect: to detect heteronuclear J-couplings, simultaneous $\pi$-pulses over H and another isotope are delivered. The central channel pictures the RF that must be applied to H nuclei, i.e. an initial $(\pi/2)_y$-pulse, followed by an alternation of $(\pi)_x$-pulses and two full Rabi oscillations along $Y$ for detection, with winding-unwinding mechanism for robustness as originally proposed in~\cite{munuera2023high} and further developed in~\cite{daly2023nutation}. The resultant magnetization of hydrogen nuclei, $\vec{M} \propto \sum_{i}^{N_H} \left[ \langle S_i^x \rangle , \langle S_i^y \rangle, \langle S_i^z \rangle\right]$(see Appendix \ref{app:samp_to_nv}), at different instants is represented with red arrows in the spheres: dashed arrows indicate the magnetization at the beginning of each protocol-step (pulses or free evolutions), while the solid arrow indicates the magnetization at the end of the step. The evolution of the magnetization outside the xy-plane (induced by RF drivings) is depicted with blue arrows indicating the magnetization's trajectory. During the detection stage, the $Y\bar{Y}$ pulse induces a magnetic field  ($B_z$) on the NV ensemble along the NV axis that precess at a controllable speed depending on the RF Rabi frequency. The amplitude of $B_z$ at the $n$-th detection period is proportional to the amplitude of the red solid arrow (i.e. to the magnetization at the end of the $n$-th encoding period). The $Y\bar{Y}$ pulses bring the magnetization back to the state at the start of the detection stage as they induce two $2\pi$ rotations. $B_z$ is captured by the NV ensemble via a MW control sequence (e.g., the XY4 in the lower channel), while green pulses indicate the standard initialization and readout of the NVs.}
\label{sequence}
\end{figure*}

\section{Results and Discussion}

\subsection{Molecular target Hamiltonian}

To exemplify the method we consider an isotropic liquid sample~\cite{comment} containing a target molecular ensemble with  two types of atoms, i.e. each molecule comprises $N_\text{H}$ hydrogens (H from now on) and a number $N_\text{A}$ of nuclei of the ``A" species. The resulting Hamiltonian is

\begin{align}
H/\hbar &= \omega_\text{H}\sum_{i=1}^{N_\text{H}}S_{i}^{z}\:+\:\sum_{i=1}^{N_\text{H}}\delta_{i}^{\text{H}}\:S_{i}^{z} + \sum_{i<j}^{N_\text{H}}J_{i,j}^\text{H}\:\vec{S}_i\cdot\vec{S}_j\nonumber\\
&+\omega_\text{A}\sum_{i}^{N_\text{A}}\:I_i^{z}\:+\:\sum_{i=1}^{N_\text{A}}\delta_i^{\text{A}}\:I_{i}^{z} + \sum_{i<j}^{N_\text{A}}J_{i,j}^\text{A}\:\vec{I}_i\cdot\vec{I}_j\nonumber\\
&+\sum_{i}^{N_\text{H}}\sum_{j}^{N_\text{A}}J_{i,j}^{\text{H}-\text{A}}\:\vec{S}_i\cdot\vec{I}_j\nonumber\\
&+ \sum_i^{N_\text{H}}\Omega_i^{\text{H}}(t)S_i^{x}\cos{\left(\omega_1\:t\right)}+\sum_i^{N_\text{A}}\Omega_i^\text{A}(t) I_i^{x} \cos{(\omega_2\:t)},
\label{eq:SfullHamiltonian}
\end{align}
where we denote the spin operators of H nuclei as $S_k^{x, y, z}$, and those of the A species as $I_k^{x, y, z}$. On the first line we have: Larmor terms of the $N_{\text{H}}$ H (with $\omega_\text{H}=|\gamma_\text{H}| B_{\text{ext}}$), their chemical shifts, and the homonuclear J-couplings of magnitude $J_{i,j}^{\rm{H}}$.  The terms on the second line are the equivalent interactions for the A species in the target molecule. The third line comprises heteronuclear J-couplings between H and A nuclei ($J_{i,j}^{\text{H}-\text{A}}$), while the last one contains the RF controls over H and A nuclear spins. Due to significant differences in gyromagnetic ratios among distinct spin species, crosstalk effects are safely neglected using the RWA, especially in high-field scenarios.
By setting the RF fields such that $\omega_1\equiv\omega_\text{H}$ and $\omega_2\equiv\omega_\text{A}$, in a rotating frame with respect to (w.r.t) Larmor terms, Eq.~(\ref{eq:SfullHamiltonian}) simplifies to  
\begin{equation}
H/\hbar = H_{shift}+\sum_{\substack{i<j \\ \text{neq}} }^{N_\text{H}}J_{i,j}^\text{H}\:S_i^z S_j^z + \sum_{i<j}^{N_\text{A}}J_{i,j}^\text{A}\:I_i^z I_j^z+\sum_{i}^{N_\text{H}}\sum_{j}^{N_\text{A}}{J}_{i,j}^{\text{H}-\text{A}}\:S_i^z I_j^z + H_c
\label{eq:IntHamiltonian1}
\end{equation}
where $H_{shift} = \sum_{i=1}^{N_\text{H}}\delta_{i}^{ \text{H}}\:S_{i}^{z} + \sum_{i=1}^{N_\text{A}}\delta_i^{\text{A}}\:I_{i}^{z} $ and $H_c = \sum_i^{N_\text{H}}\Omega_i^{\text{H}}S_i^{x}+\sum_i^{N_\text{A}}\Omega_i^{\text{A}}I_i^{x}$. For the sake of simplicity in the presentation of the method, Eq.~(\ref{eq:IntHamiltonian1}) assumes the elimination of the transversal spin operators ($I_i^{x,y}, I_j^{x,y}$, $S_i^{x,y}$ and $S_j^{x,y}$). This applies in an scenario such that $|\omega_\text{H}-\omega_i|>> J_{i,j}^{\text{H}-\text{A}}$, $|\delta^\text{H}_i-\delta^\text{H}_j|>> J^H_{i,j}$, and $|\delta^\text{A}_i-\delta^\text{A}_j|>> J^\text{A}_{i,j}$ (note, the validity of these conditions is stronger at high magnetic fields). In addition, the label ``neq" in the first summation indicates that the homonuclear J-couplings among magnetically equivalent nuclei (i.e., those with identical chemical shifts) are not included in Eq.~(\ref{eq:IntHamiltonian1}), since they do not have an impact in the dynamics as confirmed via numerical simulations. As a summary, Eq.~(\ref{eq:IntHamiltonian1}) features only ZZ type interactions and serves as a model to explain the basics of our protocol. We remark that our numerical simulations consider deviations w.r.t. Eq.~(\ref{eq:IntHamiltonian1}) due to shortcomings in the application of prior approximations and the consideration of all J-couplings throughout the entire protocol. This encompasses instances during the application of finite-width RF pulses, where deviations in the Rabi frequency are also taken into account.

\subsection{The protocol}

 J-INSECT is illustrated in Fig.~\ref{sequence}. This starts with an initial $\frac{\pi}{2}$-pulse over the H nuclei that transfer their thermal polarization to the x-axis, and continues by alternating {\it encoding} and {\it detection} stages.

During free evolutions (of length $\tau/2$) at the encoding stages, Eq.~(\ref{eq:IntHamiltonian1}) simplifies to 
\begin{equation}
H/\hbar = H_{shift}+\sum_{\substack{i<j \\ \text{neq}} }^{N_\text{H}}J_{i,j}^\text{H}\:S_i^z S_j^z +\sum_{i}^{N_\text{H}}\sum_{j}^{N_\text{A}}{J}_{i,j}^{\text{H}-\text{A}}\:S_i^z I_j^z, 
\label{eq:IntHamiltonian2}
\end{equation}
as $\sum_{i<j}^{N_\text{A}}J_{i,j}^\text{A}\:I_i^z I_j^z$ commutes with every other term in Eq.~(\ref{eq:IntHamiltonian2}) and does not impact the dynamics of H nuclei --the emitters in J-INSECT-- (check readout stage). $\pi$-pulses are delivered in the middle of the encoding stage with a twofold objective. On the one hand, they cancel $H_{shift}$ mitigating broadening caused by spatial inhomogeneities in the external magnetic field. In other words, the refocusing $\pi$-pulses lengthen the nuclear coherence time from $T_2^{*}$ to $T_2$. This enables a longer scanning duration of the nuclear sample, which proves especially advantageous for estimating J-couplings due to their relatively weak values. On the other hand, $\pi$-pulses over H (and A) enable the selective encoding of $J_{i,j}^\text{H}$ (and $J_{i,j}^{\text{H}-\text{A}}$) in the system dynamics. This becomes evident upon calculating the effective propagator at the end of the encoding stage after delivering $\pi$-pulses over H nuclei, resulting in
\begin{equation}
U_{1}(t)= \exp{\left[-i t \sum_{\substack{i<j \\ \text{neq}} }^{N_\text{H}}J_{i,j}^\text{H}\:S_i^z S_j^z \right]}.\label{eq:ZZpropagatorHOMO}
\end{equation}\label{eq:Prop1}
Alternatively, in the scenario where synchronized $\pi$-pulses are applied to H and A, resembling a double nucleus-nucleus resonance, one gets
\begin{align}
U_{2}(t)=\exp{\left[-i t \left(\sum_{\substack{i<j \\ \text{neq}} }^{N_\text{H}}J_{i,j}^\text{H}\:S_i^z S_j^z+\sum_{i}^{N_\text{H}}\sum_{j}^{N_\text{A}}J_{i,j}^{\text{H}-\text{A}}\:S_i^z I_j^z \right)\right]}.\label{eq:ZZpropagatorHETERO}
\end{align}
Besides, the influence of additional nuclear species (such as B) is negated within the dynamics due to the refocusing introduced by J-INSECT. This is, the unavoidable presence of extra nuclear spin isotopes in the target sample would not alter the system dynamics unless a $\pi$-pulse that synchronously flips the B species is introduced on purpose (see later).

Then, the initial $\frac{\pi}{2}$-pulse along the y-axis takes the thermal state of hydrogen nuclei to the x-axis (see Fig.~\ref{sequence}) while  ZZ-interactions that govern the encoding stage --i.e. those in Eqs.~(\ref{eq:ZZpropagatorHOMO}) or~(\ref{eq:ZZpropagatorHETERO})-- encode J-couplings in the dynamics of $\langle S_i^x\rangle$ and $\langle S_i^y\rangle$ of H nuclei. Importantly, owing to the initialization of the thermal polarization along the x-axis, it can be demonstrated that following the encoding stage, $\langle S_i^x\rangle$ exhibits significantly greater amplitude than $\langle S_i^y\rangle$ (approximately $10^{5}$ times larger), as detailed in Appendix \ref{app:amplitudes}. In this manner, J-INSECT is tuned to target $\langle S_i^x\rangle$ in the detection stages.

At a generic $n$-th detection stage, owing to a rotation of the H nuclei along the y-axis (see Fig.~\ref{sequence}.), the magnetic field $B_z^n$ induced by the sample over the NV ensemble is along the NV axis and reads~\cite{meriles2010imaging} (see Appendix \ref{app:samp_to_nv})
\begin{equation}\label{effectiveB}
B_z^n(t)= B_0(n\tau) \sin(\Omega^\text{H} t),
\end{equation}
where $\Omega^\text{H}$ is the Rabi frequency of the $Y$ and $\bar{Y}$ pulses (see Fig.~\ref{sequence}) while, importantly:
\begin{equation}
B_0(n\tau)\propto \frac{2}{N^\text{H}}\sum{\langle S_i^x\rangle_{t=n\tau}}.
\end{equation}
In conclusion, the subsequent magnetic field $B^n_z(t)$ has an amplitude proportional to $\langle S_i^x\rangle$ of H nuclei, and oscillates with a controllable frequency $\Omega^\text{H}$. This is, irrespective of the involved Larmor frequencies, what enables to operate at high magnetic fields. Note that, at 2 T hydrogen Larmors reach $\approx (2\pi)\times85$ MHz, posing a technically challenging task for their tracking. On the contrary, $\Omega^\text{H}$ can be tuned, for instance to tens of kHz \cite{herb2020broadband, yudilevich2023coherent}, such that the oscillations of $B^n_z(t)$ are easily followed by the NV ensemble by applying, e.g., a standard XY4 sequence. Recording the resultant phase acquisition of the NV ensemble after every detection period, leads to a direct reconstruction of the evolution of $\langle S_i^x\rangle$ of H nuclei, thus of the J-couplings encoded in $\langle S_i^x\rangle$. In this regard, it is worth noting that, unlike standard heterodyne measurement schemes, J-INSECT does not impose restrictions on the separation between consecutive measurements. This means they can be placed arbitrarily in time, provided that the elapsed time between measurements is recorded, as this is crucial for extracting chemical information. For simplicity in the protocol, they are placed equidistant, separated by $\tau$. Notice that since hydrogens are left in the same state as before the measurement (due to the rotation being a multiple of $2\pi$, the next encoding period can be applied directly, without the need to repolarize the sample).
Finally, it is important to note that the NV centers are active only during the detection periods; this implies a high spectral resolution constrained not by the sensor, but by the $T_2 (> T_2^{*})$ of H nuclei.

\subsection{Numerical results} 

We consider a picoliter sized liquid sample on top of a diamond hosting an NV ensemble at micrometer depth~\cite{glenn2018high} and a static magnetic field of $B_{\text{ext}}=2$ T aligned with the NV axis. 

\begin{figure*}[t]
\includegraphics[width=1 \linewidth]{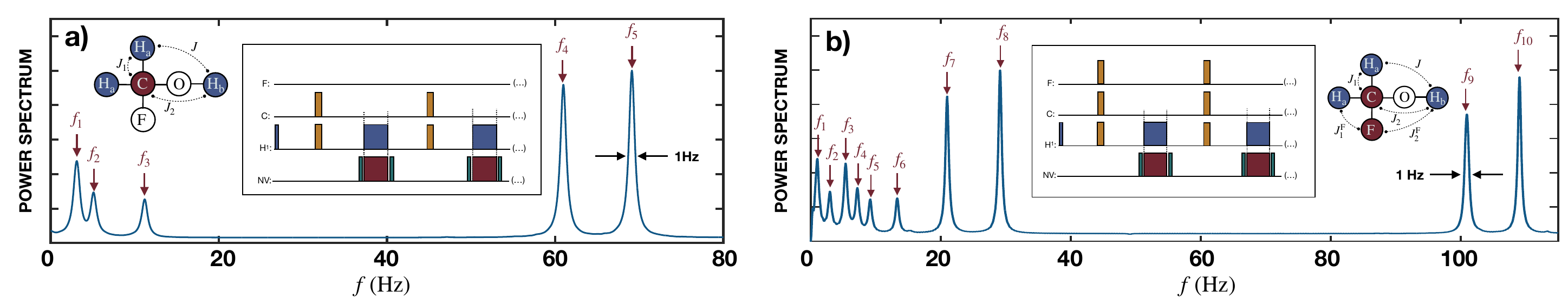}
\caption{{\bf Simulated spectra for the fluoromethanol molecule}.  Obtained from 1000 repetitions of the protocol with 600 \emph{encoding} + \emph{detection} stages. Due to the dephasing time $T_2=0.6$~s, the FWHM is approximately 1 Hz.  Within each figure, there is an illustrative depiction of the molecule, highlighting the targeted J-couplings. Colored nuclei signify the species upon which $\pi$ pulses are applied. Additionally, each figure includes a boxed scheme detailing the RF pulse train applied on each case. (a) Targeting only H and $^{13}$C, five peaks arise, which encode $J$, $J_1$ and $J_2$. b) Targeting the H, the $^{13}$C and the F, ten peaks arise, encoding: $J$, $J_1$, $J_2$, $J_1^{\rm{F}}$ and $J_2^{\rm{F}}$.} 
\label{fig:spectra}
\end{figure*}

To exemplify J-INSECT we address a sample consisting on fluoromethanol molecules ($\rm{CH_2FOH}$)  that contains five spin-1/2 nuclei of three different species: H, carbon and fluorine ($^{13}$C and F from now on, note oxygen has no spin) in a thermal state, see scheme of the molecule on inset of Fig.~\ref{fig:spectra}. Our protocol enables accessing heteronuclear J-couplings among H and the $^{13}$C nucleus, as well as with the F. Moreover, it gives access to the homonuclear coupling between non-equivalent H (the two H nuclei in Fig.~\ref{fig:spectra} labeled as H$_\text{a}$ are magnetically equivalent, whereas H$_\text{b}$ is magnetically unequivalent with respect to the former). We choose values falling within the typical range found in the literature \cite{levitt2013spin}, namely: $J_{\rm{\text{H}_\text{a}\text{H}_\text{b}}}\equiv J=(2\pi)\times8\:\rm{Hz}$, $J_{\rm{\text{H}_\text{a}C}}\equiv J_1=(2\pi)\times130\:\rm{Hz}$, $J_{\rm{H_bC}}\equiv J_2=(2\pi)\times6\:\rm{Hz}$, while interactions with the F nuclei are $J_{\rm{H_aF}}\equiv J^F_1=(2\pi)\times80\:\rm{Hz}$ and $J_{\rm{H_bF}}\equiv J^F_2=(2\pi)\times4\:\rm{Hz}$, whereas $J_{CF}=(2\pi)\times160$. Furthermore, for a magnetic field $B_{\text{ext}}=2$ T, the chemical shifts are $\delta_{\rm{H_a}}=(2\pi)\times512 \rm{Hz}$, $\delta_{\rm{H_b}}=(2\pi)\times236 \rm{Hz}$, $\delta_{\rm{C}}=(2\pi)\times85 \rm{Hz}$ and $\delta_{\rm{F}}=(2\pi)\times450 \rm{Hz}$.

Our numerical simulations start with Hamiltonian~(\ref{eq:SfullHamiltonian}) in an interaction picture w.r.t the Larmor terms $\omega_H\sum_{i=1}^{N_\text{H}}S_{i}^{z} + \omega_\text{A}\sum_{i}^{N_\text{A}}I_i^{z}$. In this manner, the only approximation made is simplifying the heteronuclear coupling term $\sum_{i}^{N_\text{H}}\sum_{j}^{N_\text{A}}J_{i,j}^{\text{H}-\text{A}}\:\vec{S}_i\cdot\vec{I}_j$ in Eq.~(\ref{eq:SfullHamiltonian})  to only ZZ interactions  (i.e. to $\sum_{i}^{N_\text{H}}\sum_{j}^{N_\text{A}}J_{i,j}^{\text{H}-\text{A}}\: S_i^z I_j^z$, like in Eq.~(\ref{eq:IntHamiltonian1})). This can be done since $|\omega_\text{H}-\omega_i|\sim (2\pi)\times10 \text{ MHz}>> J_{i,j}^{\text{H}-\text{A}}\sim (2\pi)\times100\text{ Hz}$, thus, the rotating wave approximation safely applies. The reader can find the full simulated Hamiltonian in Appendix \ref{app:samp_to_nv}.

In order to maintain the Hamiltonian close to Eq.~(\ref{eq:IntHamiltonian1}) the condition $\tau > 1/(|\delta_{\rm{\text{H}_\text{a}}}-\delta_{\rm{\text{H}_\text{a}}}|)$ must hold, such that the homonuclear term of H spins approaches to a ZZ interaction. The previous condition imposes a lower bound on $\tau$, limiting the amount of measurements that can be carried out for a given experimental time. In our case, we set $\tau = 1.2/(|\delta_{\rm{\text{H}_\text{a}}}-\delta_{\rm{\text{H}_\text{a}}}|)=4.3$ ms (note later we explore the J-INSECT performance with a larger number of measurements, i.e., departing from this condition). In addition, the amplitude of the RF targeting H is $\Omega^{\text{H}}=(2\pi)\times 50$ kHz while, for robustness purposes, $\pi$-pulses over H are implemented via a CORPSE architecture~\cite{Cummins2003tackling} which reduce the impact of detunings during pulse execution. We observe no discernible impact on the system dynamics when standard top-hat pulses are applied to $^{13}$C and F, thus we exclusively deploy CORPSE pulses over Hs. This results in $t_{\pi}^{CORPSE}=43.3 \ \mu$s, $t_{\pi}^{\rm{C}}=39.7 \ \mu$s and $t_{\pi}^{\rm{F}}=10.6 \ \mu$s. Our simulations firstly consider 600 measurements (this is, 600 encoding and detection stages in a single experimental run) hence, the duration J-INSECT per experimental run is $T_{total}=600\:\left(\tau+t_{\pi}^{CORPSE}+2/\Omega^{\rm{\text{H}}}\right)\approx2.66\:\rm{s}$, where the term $2/\Omega^{\rm{\text{H}}}$ corresponds to the duration of two full RF Rabi oscillations during the readout stage.

Regarding error and imperfections, our numerical simulations comprise: (1) Always on chemical shifts and J-couplings, (2) RF Cross-talk effects during RF irradiation, (3) Errors on the amplitude of the RF drivings (these are simulated through an Ornstein-Uhlenbeck process \cite{uhlenbeck1930on,wang1945on,gillespie1996exact} with a 1\% amplitude shift and a noise correlation time of 1 ms) and (4) A nuclear dephasing time $T_2=0.6\:\rm{s}$~\cite{glenn2018high}, incorporated via an exponential decay in the field emitted by the sample.

We consider two cases: When J-INSECT targets H and $^{13}$C nuclei (Case 1) we get the spectrum in Fig.~\ref{fig:spectra}.a. Note that in this case, F does not actively participate, since we do not apply $\pi$-pulses over it, although it is included in our numerical model. Case 1 exhibits five resonances at different frequencies $f_i$ encoding heteronuclear J-couplings between  H and $^{13}$C and homonuclear J-coupling between H nuclei. The assignment of J-couplings based on the obtained $f_i$ can be accomplished either by analytically solving the evolution in Eq.~(\ref{eq:ZZpropagatorHETERO}) (see Appendix \ref{app:fluoromethanol}) or by employing standard spin multiplicity techniques \cite{multiplicity}. Conversely, simultaneous targeting of the F nuclei with an additional $\pi$-pulse (Case 2) results in the emergence of ten resonances, as illustrated in Fig.~\ref{fig:spectra}.b. Note, in this second scenario the heteronuclear coupling between the F and H nuclei is also encoded in $B_0(n\tau)$. The resulting estimations for the J-couplings in Cases 1 and 2 are collected in Table~\ref{table:values}. 

\begin{table}[h]
\begin{tabular}{ | c | c | c | c | c | c |} 
  \hline
   & $\langle J\rangle$ & $\langle J_1\rangle$ & $\langle J_2\rangle$ & $\langle J_1^{\rm{F}}\rangle$ & $\langle J_1^{\rm{F}}\rangle$ \\ 
  \hline
  Case 1 & $8.1\pm0.5$ & $130.1\pm0.8$ & $6.1\pm0.9$ & - & - \\ 
  \hline
  Case 2 & $8.1\pm0.5$ & $130.1\pm0.9$ & $6\pm1$ & $80.0\pm0.9$ & $4\pm1$ \\ 
  \hline
\end{tabular}
\caption{J-coupling values obtained from the spectra shown in Fig.\ref{fig:spectra}.a (first row) and Fig.\ref{fig:spectra}.b (second row). Values and their uncertainty are given in Hertz.}
\label{table:values}
\end{table}

In summary, J-INSECT effectively discloses J-couplings with large spectral resolution resulting in a full-width-at-half-maximum (FWHM) of, at most, 1 Hz, corresponding to the chosen value of $T_2=0.6$ s. Regarding sensitivities, the use of elevated magnetic fields allows achieving a signal-to-noise ratio (SNR) of 30 with approximately 18,000 repetitions of the protocol (estimation carried out for Case 1 with photon statistics of NV experiments, assuming  a 7\% of luminescence contrast.  Additional details concerning Case 1 and the development of Case 2, which yields similar results, are available in Appendix \ref{app:readout}). This implies an experimental time of $\sim$13h.  Hence, J-INSECT facilitates a high-fidelity retrieval of J-couplings within a reasonable signal-collecting time, with the potential to further reduce the experimental duration by incorporating complementary methods, such as prior hyperpolarization of the sample reported in Refs.~\cite{bucher2020hyperpolarisation,arunkumar2021micron}, or advanced readout techniques such as repetitive readout~\cite{photoncount,repetitivereadout}. In terms of sensitivity, previous works using quantum heterodyne measurements at moderate magnetic fields ($88$ mT) \cite{glenn2018high} report sensitivities in the range of $\sim 25-75{\rm pT\cdot Hz^{-1/2}}$. Since our work operates at high magnetic fields, which enhance polarisation by a factor of $\sim 30$, we anticipate a sensitivity of approximately $\sim 1-5{\rm pT\cdot Hz^{-1/2}}$. Additionally, our approach extends the coherence time of the sample from $T_2^*$ to $T_2$, which will further reduce the sensitivity.

An interesting final aside to consider is enhancing the rate of measurements, as an strategy to extra reduce experimental time. To exemplify the procedure we inspect the simpler Case 1. Enhancing the measurement rate implies reducing $\tau$ in the encoding stages, hence departing from the condition $\tau > 1/(|\delta_{\rm{H_a}}-\delta_{\rm{H_a}}|)$. In this scenario the spectrum is no longer given by the usual multiplicity rules (i.e. the homonuclear terms, H-H, the interaction Hamiltonian Eq.\ref{eq:IntHamiltonian1} are no longer ZZ interactions), thus we estimate the values of the J-couplings by performing Bayesian inference. In particular, we choose to extend the number of measurements on each run (i.e. the number of detection stages) from 600 to 6553 and reduce the overall experimental time four times (from $\sim$ 13 h to $\sim$ 3h 30'). In this case, the inference gives $\langle J\rangle=8.2\pm0.5$, $\langle J_1\rangle=130.1\pm0.6$ and $\langle J_2 \rangle=5.7\pm0.4$ (see Appendix \ref{app:beyond}). 

\section{Conclusions} 

We introduced J-INSECT, a protocol that enables the detection of homonuclear and heteronuclear J-couplings with NV ensembles at high magnetic fields. J-INSECT incorporates heterodyne techniques and exerts noise filtering, enhancing the sample's coherence thus enabling highly precise estimations of J-couplings limited by $T_2 > T_2^{*}$. We explore J-INSECT's performance in two  scenarios, considering both reduced and enhanced data harvesting. We assess the influence of different noise sources, including the persistent presence of always-on J-couplings, showcasing the method's feasibility in experimental conditions.

\section{data availability}
The complete dataset supporting the findings of this work is accessible. On the one hand, secondary data reused and/or analyzed from other studies is specified along the text, in case any doubts or inquires, please contact the corresponding authors at jcasanovamar@gmail.com. On the other hand, original data generated in this study (via numerical simulations) are available upon request, kindly direct any further correspondence to the aforementioned email address.

\section{code availability}
The source code and associated materials used in this research are available upon request. For access to the code, inquiries, or additional information, please contact the corresponding authors at jcasanovamar@gmail.com. We are committed to facilitating transparency and collaboration in our research and will provide assistance and code access as needed.

\section{Acknowledgements}

P. A. B. and A. B. U. acknowledge the financial support of the IKUR STRATEGY (IKUR-IKA-23/22) and (IKUR-IKA-23/04), respectively. C. M. J. acknowledges the predoctoral MICINN Grant No. PRE2019-088519. J. C. acknowledges the Ram\'{o}n y Cajal (RYC2018-025197-I) research fellowship. Authors acknowledge the Quench project that has received funding from the European Union's Horizon Europe -- The EU Research and Innovation Programme under grant agreement No 101135742, the financial support from Spanish Government via the Nanoscale NMR and complex systems (PID2021-126694NB-C21) project, the ELKARTEK project Dispositivos en Tecnolog\'{i}as Cu\'{a}nticas (KK-2022/00062), and the Basque Government grant IT1470-22.

\appendix

\section{Spin dynamics of the sample}
In this section, we provide further details regarding the nuclear spin dynamics in the sample.

\subsection{Amplitudes of the $\langle\sigma_i^x\rangle$ and $\langle\sigma_i^y\rangle$ components}\label{app:amplitudes}

We consider an evolution governed by the next propagator  (note this is similar to those in Eqs.~(4,5) in the main text)
\begin{equation}
U_t = \Pi_{i,j}e^{-i\frac{\phi_{i,j}}{2} \sigma_i^z\sigma_j^z},
\end{equation}
over an initial state such that 
\begin{equation}
(\rho_1^x ... \rho_{k}^x)(\rho_{k+1}^z ... \rho_{N}^z),
\end{equation}
where $\rho_j^{x,z}$ are thermal states polarized along $x$ and $z$ directions. This is, $\rho_j^{z} = \frac{1}{2}I + \frac{B_{\rm n}}{4} \sigma_z$ while $\rho_j^{x} = e^{i\frac{\pi}{4} \sigma_j^y}\rho_j^{z} e^{-i\frac{\pi}{4} \sigma_j^y} = \frac{1}{2}I - \frac{B_{\rm H}}{4} \sigma_x$, with $B_H=\frac{\hbar \gamma_h B_{\rm{ext}}}{K_b T}$ and $B_n=\frac{\hbar \gamma_n B_{\rm{ext}}}{K_b T}$ being small quantities at room temperature (on the order of $10^{-5}$ for a magnetic field of 2 T). In the previous expressions, $B_{\rm ext}=2\:\rm{T}$ is the external magnetic field, $\hbar=1.054\cdot 10^{-34}\:\rm{J\cdot s}$ is the Planck constant divided by $2\pi$, $\gamma_H=(2\pi)\times42.57\: \rm{MHz/T}$ is the gyromagnetic ratio of the H and $\gamma_n$ the corresponding one of other nuclei, $K_B=1.38\cdot10^{-23}\:\rm{J/K}$ the Boltzmann constant, and the temperature $T=300$ K. Here we differentiate among thermal states corresponding to the hydrogen (H) nuclei, i.e. the nuclear spin which is rotated to the X axis at the start of the protocol (denoted by $\rho_j^x$), and the other generic nuclear spin, which is represented with $\rho_j^z$.
 
On the one hand, one can demonstrate that

\begin{align}
\langle \sigma_1^x \rangle &= {\rm Tr}[U_t (\rho_1^x ... \rho_{k}^x)(\rho_1^z ... \rho_{k}^z)U^\dag_t  \sigma_1^x]\nonumber\\
&= {\rm Tr_{\neq 1}}[\langle e|\rho_1^x| g\rangle ... \rho_{k}^x \rho_{k+1}^z ... \rho_{N}^z \Pi_{j>1}e^{-i\phi_{1,j} \sigma_j^z} \nonumber\\
&+ \langle g|\rho_1^x| e\rangle ... \rho_{k}^x \rho_{k+1}^z ... \rho_{N}^z \Pi_{j>1}e^{i \phi_{1,j} \sigma_j^z}],
\label{sigmax}
\end{align}

where $\rm{Tr}_{\neq 1}[...]$ represents the trace over all spins excepting the first one. Using $\langle e|\rho_1^x| g\rangle = \langle g|\rho_1^x| e\rangle = - \frac{B_{\rm H}}{4}$ one reaches 

\begin{eqnarray}
\langle \sigma_1^x \rangle &=  -\frac{B_{\rm H}}{4}\Pi_{j=1}^k \cos{\phi_{1,j}} \ \Pi_{j=k+1}^N \bigg(\cos{\phi_{1,j}} - \frac{iB_{\rm n}}{2}\sin{\phi_{1,j}} \bigg) \nonumber\\
&- \frac{B_{\rm H}}{4}\Pi_{j=1}^k\cos{\phi_{1,j}} \ \Pi_{j=k+1}^N \bigg(\cos{\phi_{1,j}} + \frac{iB_{\rm n}}{2}\sin{\phi_{1,j}} \bigg).
\end{eqnarray}

On the other hand, one can demonstrate that $\langle \sigma^y_1 \rangle$ leads to 
\begin{eqnarray}
\langle \sigma_1^y \rangle &=- \frac{iB_{\rm H}}{4}\Pi_{j=1}^k\cos{\phi_{1,j}} \ \Pi_{j=k+1}^N \bigg(\cos{\phi_{1,j}} - \frac{iB_{\rm n}}{2}\sin{\phi_{1,j}} \bigg) \nonumber\\
&+ \frac{iB_{\rm H}}{4}\Pi_{j=1}^k\cos{\phi_{1,j}} \ \Pi_{j=k+1}^N \bigg(\cos{\phi_{1,j}} + \frac{iB_{\rm n}}{2}\sin{\phi_{1,j}} \bigg).
\end{eqnarray}
Note the above expressions hold for any H, thus for any quantity $\langle \sigma_i^x \rangle$ and $\langle \sigma_i^y \rangle$ that involves a H spin in the target sample.

Notice how the leading order for each expectation value differs by five order of magnitude: $\langle\sigma_1^x\rangle\propto\mathcal{O}\left(B_H\right)$ and $\langle\sigma_1^y\rangle\propto\mathcal{O}\left(B_H B_n\right)$.  Thus, $\langle\sigma_i^x\rangle$ is $\sim10^5$ times greater than $\langle\sigma_i^y\rangle$.

\subsection{Fluoromethanol resonances}\label{app:fluoromethanol}
We compute the resonance positions for the specific case of the fluoromethanol molecule. For the sake of simplicity we consider a polarised H, and the other nuclei in the maximally mixed state, this is $\rho_i^x=\frac{1}{2}(I+\sigma_i^x)$ for H and $\rho_i^m=\frac{1}{2}I$ for other nuclei. Note, these considerations only alter the amplitude of the signal, not the resonance position. However, the results in the main text consider thermal states for all nuclei in the fluoromethanol molecule. Thus, following Eq.~(\ref{sigmax}), we have 
\begin{eqnarray}
\langle S_i^x \rangle = \frac{\langle\sigma_i^x\rangle}{2}= \frac{1}{2}\Pi_{j=1}^N \cos{\phi_{i,j}}.\nonumber
\end{eqnarray}
where $\phi_{1,i}$ corresponds to the phase accumulated due to the interaction between the i-th and j-th spins during a time t, that is $\phi_{i,j}(\text{t})=\frac{J_{i,j}}{2}\text{t}$.

When H nuclei and $^{13}$C are targeted in the fluoromethanol molecule (corresponding to Case 1 in the main text), the resulting signal is 

\begin{eqnarray}
 \langle\sum_{i=1}^{3}{S^x_i}\rangle (\text{t})=\frac{1}{6}\cos{\left(\frac{J}{2}\text{t}\right)}\left[2\cos{\left(\frac{J_1}{2}\text{t}\right)}+\cos{\left(\frac{J}{2}\text{t}\right)}\cos{\left(\frac{J_2}{2}\text{t}\right)}\right],\nonumber
\end{eqnarray}
which leads to the resonance peaks in Fig.2.a of the main text, $f_1=\pm J_2/2$, $f_2= \pm(J_2 -2 J)/2$, $f_3= \pm (J_2 +2 J)/2$, $f_4=  \pm(J_1-J)/2$, and $f_5= \pm (J_1+J)$. 

In the case where the fluorine (F) is also targeted with $\pi$-pulses (corresponding to Case 2 in the main text), one finds

\begin{align}
 \langle\sum_{i=1}^{3}{S^x_i}\rangle (\text{t}) =\frac{1}{6}\cos{\left(\frac{J}{2}\text{t}\right)}&\left[2\cos{\left(\frac{J_1}{2}\text{t}\right)}\cos{\left(\frac{J_1^{\rm{F}}}{2}\text{t}\right)}\right.\nonumber\\
 &\left. +\cos{\left(\frac{J}{2}\text{t}\right)}\cos{\left(\frac{J_2}{2}\text{t}\right)}\cos{\left(\frac{J_2^{\rm{F}}}{2}\text{t}\right)}\right],\nonumber
\end{align}

with the corresponding resonances depicted in Fig.2.b of the main text. These are: $f_1= \pm \left(J_2-J_2^{\rm{F}}\right)/2$, $f_{2}=\pm \left(+2J-J_2-J_2^{\rm{F}}\right)/2$, $f_3=\pm \left(J_2+J_2^{\rm{F}}\right)/2$, $f_4=\pm \left(2J+J_2^{\rm{F}}-J_2\right)/2$, $f_5=\pm \left(2J+J_2-J_2^{\rm{F}}\right)/2$, $f_6=\pm \left(2J+J_1+J_1^{\rm{F}}\right)/2$, $f_7=\pm \left(J_1-J_1^{\rm{F}}-J\right)/2$,  $f_8=\pm \left(J+J_1-J_1^{\rm{F}}\right)/2$, $f_9=\pm \left(J_1+J_1^{\rm{F}}-J\right)/2$ and $f_{10}=\pm \left(J+J_1+J_1^{\rm{F}}\right)/2$. 

Alternatively, the resonant frequencies can be obtained via standard NMR multiplicity rules~\cite{multiplicity}. This is, when a nuclei A is coupled with $n_x$ spin-$\frac{1}{2}$ nuclei of an isotope X, the A signal splits into $n_x+1$ lines. If A is also coupled to an element different to X (lets say Y) this induces additional $n_y+1$ splittings in each previous peak. 

In our case, since the broadening is mitigated, all J-coupling peaks tend to be centered at 0 Hz. Thus, when H and $^{13}$C are targeted, the signal of $H_a$ (recall the $H_a$ and $H_b$ are magnetically inequivalent hydrogen groups in the molecule, refer to Fig. 2 of the main text) is split into $2\times2=4$ peaks, due to couplings with $H_b$ and the $^{13}$C, while the signal of $H_b$  is split into $3\times2=6$ peaks, 3 due to the pair of equivalent $H_a$, and 2 due to the $^{13}$C. Then, we are left with $6+4=10$ peaks around 0 Hz. If J coupling interactions with F are also excited, the signal of $H_a$ is split into $2\times2\times2=8$ (2 per every other element), while the signal of $H_b$  splits into $3\times2\times2=12$ peaks, leading to a total of 20 peaks, see, e.g.~\cite{abragam1961principles} for more information.

\section{From the sample to the NV centers}\label{app:samp_to_nv}

The oscillation of hydrogens in the detection stage of J-INSECT impacts on the NV ensemble (which is considered at a micrometer depth from diamond surface) as a classical magnetic field. In this respect, see e.g.~\cite{glenn2018high,meriles2010imaging}). In the high field regime, the magnetic signal induced by the sample in the orthogonal directions (i.e. x and y) with respect to the external magnetic field (oriented with the NV axis, i.e. $z$) can be neglected as it rapidly oscillate at a speed $\gamma_N B_{z}$ with $B_z$ being several Tesla. The resultant field emanated from the sample during a n-th detection period ($n \in \mathbb{N}  $) can be well approximated by~\cite{glenn2018high, meriles2010imaging} 
\begin{align}
B_z(t)= B_0^n \sin(\Omega^H t)
\label{eq:app:SampleSignal}
\end{align}

where $\Omega^H=\gamma_H B_{RF}$ is the amplitude of the RF driving. On the other hand, the amplitude of the magnetic field is
\begin{equation}
B_0^n=\frac{(2\pi)^2(\hbar\:\gamma_H)^2\mu_0\:\rho_H\: B_z F_3 }{16\pi K_B T} M_x
\end{equation}
where $M_x$ is the x component of the normalized magnetisation, defined as $\vec{M}=\frac{2}{N^H} \langle \sum_i{\vec{S}_i}\rangle (n\tau)$. Note that the quantity  $\langle\sum_i{\vec{S}_i}\rangle(n\tau)$ is computed via numerically evolving the next molecular Hamiltonian:
 \begin {align}
H/\hbar= &\delta^{H_A} (S_1^z+S_2^z)+\delta^{H_B}S_3^z+\delta^C I^z +\delta^F P^z \nonumber\\
&+ J_{H_aH_a} \vec{S_1}\cdot \vec{S_2} + J_{H_aH_b} (\vec{S_1}\cdot \vec{S_3}+\vec{S_2}\cdot \vec{S_3}) \nonumber\\
     &+ J_{H_aC} (S_1^z+S_2^z)I^z+ J_{H_bC} S_3^z I^z \nonumber\\
     &+ J_{H_aF} (S_1^z+S_2^z)P^z+ J_{H_bF} S_3^z P^z + J_{CF} I^z P^z
 \end{align}
 
In addition, $\hbar=1.054\cdot 10^{-34}\:\rm{J\cdot s}$ is the Planck constant divided by $2\pi$, $\gamma_H=(2\pi)\times42.57\: \rm{MHz/T}$ is the gyromagnetic ratio of the H, $\mu_0=4\pi\cdot10^{-7}\:\rm{H/m}$ the magnetic permeability of free space, $\rho_H\approx 6.6\times10^{28}\: m^{-3}$ the H density of the sample, $B_z=2\:\rm{T}$ the external magnetic field, $K_B=1.38\cdot10^{-23}\:\rm{J/K}$ the Boltzmann constant, $T=300K$ the temperature and $F_3=4.1$ characterizes the geometry of the sample~\cite{glenn2018high}.

The Hamiltonian of a generic NV in the ensemble under the influence of $B_z(t)$ is 
\begin{align}
H_{NV}=\frac{\sigma_z\:\gamma_e}{2}B_z(t) +H_c,
\label{eq:app:NVHamiltonian}
\end{align}
where $H_c$ accounts for the MW driving on the NV ensemble. In particular, applying a XY4 control sequence during the detection stage over the NV-ensemble with $t_{RF}=2 \frac{1}{\Omega^H}$ and the spacing between the MW pulses $T_{MW}=t_{RF}/4$, yields:
\begin{align}
    \langle\sigma_y\rangle_{NV}\approx\frac{2\gamma_e\:t_{RF}}{\pi}B_0^n,
    \label{eq:app:NV signal}
\end{align}
provided that $\frac{2\:\gamma_e t_{RF}}{\pi}\:B_0^n<<1$.

\section{READOUT CONSIDERATIONS}\label{app:readout}
In order to evaluate the impact of the measurement rate of the protocol under the high external magnetic field paradigm, we make an estimation of the required signal averaging time leading to a SNR of 30.

In \cite{photoncount} they record the mean number of photon-count per measurement when a single NV is in $\ket{0}$, that is $n_0=0.016$, and a photon count difference between the ground and excited spin state of $n_0-n_1$=0.005, yielding a $30\%$ of luminescence contrast between both states. In this case, since we are working with NV ensembles, we considered a fluorescence contrast of $7\%$, which was achieved by lowering the photon emission $n_0$. Employing the latter photon statistics, and taking as a basis the signal with no readout error (in particular, the inverse Fourier transform of the spectra in Fig. 2.a, note that this signal carries the rest of errors described in the main text), we have reproduced an experimental outcome. 

In particular, we assume a configuration similar to that in Ref.~\cite{glenn2018high}. This is an NV center density of $0.8\times 10^{23}\text{m}^{3}$, a beam diameter of $20 \ \mu$m, and a $10 \ \mu$m thick layer of NV centers  sensitive to the thermal spin signal. This leads to a sensor containing $\approx 2.5\times10^8$ active NV centers. The protocol is repeated $18000$ times. This is, we  average the emitted signal  across $18000$ repetitions of the protocol, which leads to a duration of $18000\times2.66\text{s}\approx\text{13\text{ h }}20'$.

\begin{figure}[]
\includegraphics[width= 1 \linewidth]{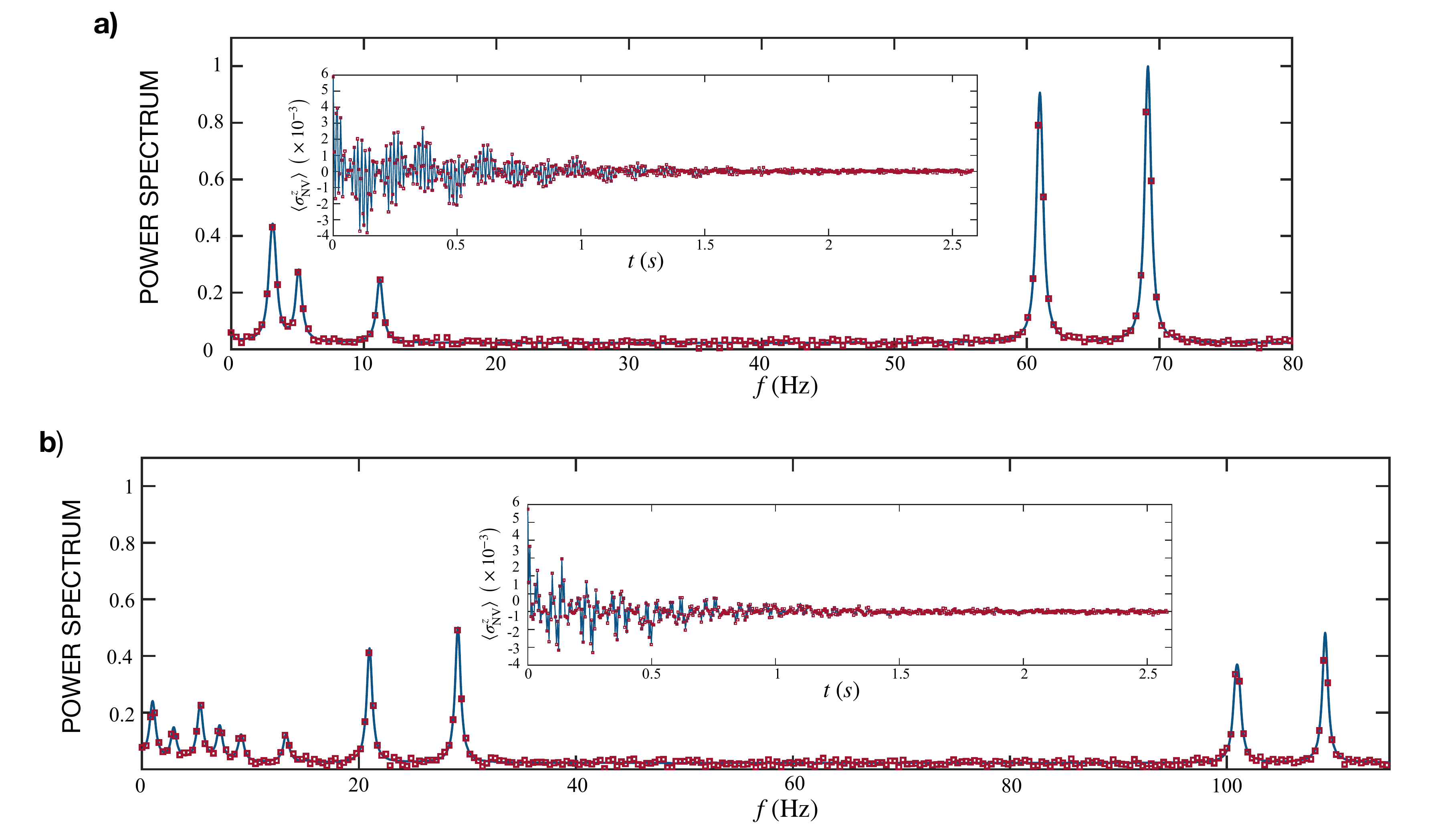}
\caption{Simulating an experiment. Figure a) corresponds to Case 1, while b) holds for Case 2. Insets: in blue are the signals without considering luminescence issues, while red squares are the outcome that results from averaging $18000$ repetitions of the respective protocol with $2.5\times10^8$ NV centers assuming a contrast of 7\%. Main figures: In blue are the spectra that correspond to the signals without luminescence issues, while the red squares are the FT of averaged signal in the inset (thus, carrying readout noise). Both spectra are normalized by a common factor, ensuring that the maximum peak in Case 1 has amplitude 1. In Case 1, referencing the amplitude of the smallest peak, the corresponding SNR is just above 30. In Case 2 under the same reference criterion, it is slightly below 15.}\label{app:fig:exp}
\end{figure}

Specifically, in Case 1, the noise in the resulting averaged signal has a standard deviation of approximately  $6\times 10^{-5}$  while the signal's amplitude is $\langle \sigma_{NV}^z \rangle\sim 1\times 10^{-3}$, see inset of  Fig.~\ref{app:fig:exp}.a. This leads to a SNR of about 30 in the spectrum by taking the the height of the smallest peak. For Case 2, the noise standard deviation is similar. However, the SNR is approximately 15. This is because even though both signals (from Case 1 and Case 2) have similar amplitudes, the spectrum of Case 2 exhibits more peaks which results in peaks of smaller amplitude as it can be seen in Fig.~\ref{app:fig:exp} for clarification.

In conclusion, the thermal polarization induced by the high external magnetic field is enough to balance out the relatively low measurement rate, yielding a detectable signal (thus of clean J-coupling spectra) under reasonable experimental time in both cases. 

\section{Beyond $\tau > 1.2/(|\delta_{\rm{H_a}}-\delta_{\rm{H_a}}|)$}\label{app:beyond}
\begin{figure*}[]
\includegraphics[width=1 \linewidth]{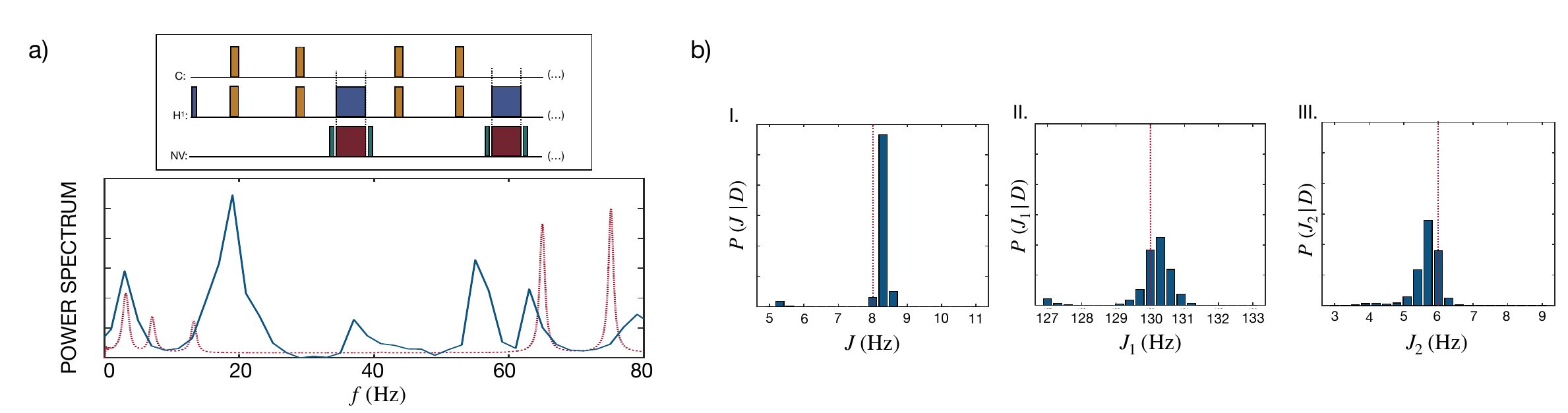}
\caption{ a) Top, scheme of the RF sequence employed for the high measurement-rate protocol (with $(\pi)_x$ and $(\pi)_{-x}$ pulses altogether with CORPSE structures over Hs during encoding for enhanced robustness) and the corresponding spectrum (bottom). This shows the result obtained with 6553  measurements (solid blue) while, for comparison, we include the spectrum obtained with 600 measurements (dashed red), which corresponds to the same spectrum shown in Fig. 2.a of the main text. b) Posterior distributions for each scanned parameter via Bayesian inference. I: Posterior distribution corresponding to J, II: to $\rm J_1$ and III: to $\rm J_2$. The exact value of each parameter is marked by a vertical dotted line in I, II, and III. The uncertainties are the standard deviation $\sigma_{\theta}$, given by $\sigma_{\theta}^2=\langle(\theta-\langle\theta\rangle)^2\rangle=\int(\theta-\langle\theta\rangle)^2\:P(\theta|\rm{D})\:d\theta$, where $\theta$ stands for J, $\rm J_1$ and $\rm J_2$.} 
\label{app:fig:distr}
\end{figure*}
To increase the number of measurements on each experimental run, we reduce $\tau$ in encoding stages from $\tau=4.3\text{ ms}$ to $\tau=75\:\mu s$ and perform 6553 measurements instead of 600.  We consider imperfect pulses (with 1\% error on the amplitude driving) as well as always-on J-couplings and readout errors (see previous section). This increase in the number of measurements leads to a reduction on the total experimental repetitions to be performed. In this case the protocol includes an extra ($\pi$)-pulse during the encoding period to further reduce the impact of imperfect drivings (see the scheme of the pulse sequence in Fig.~\ref{app:fig:distr}).

On the other hand, the reduction on $\tau$ leads to an intricate spectrum without clear resonance peaks, see Fig.~\ref{app:fig:distr}.a). Thus, to infer the values of J-couplings we perform Bayesian analysis. As an input to the Bayesian inference we employ the previously described noisy signal with imperfect pulses and readout errors. Regarding the parameter set, as the prior knowledge we assume a uniform distribution. To sum up, via Bayes inference, simplicity of the spectra can be traded for shorter experimental times. 

As a figure of merit, we simulate the resulting signal performing four times less averages (thus a four-times shorter experiment). This results into a readout error twice greater. The posterior distributions are depicted in Fig \ref{app:fig:distr}.b, from which the following values and uncertainties arise (respectively): $\langle J\rangle=8.2\pm0.5$ Hz, $\langle J_1\rangle=130.1\pm0.6$ Hz and $\langle J_2 \rangle=5.7\pm0.4$ Hz.


\begin{thebibliography}{99}
\bibitem{comment} The full form of the J-coupling is a tensor, although in an isotropic liquid --this is, in common liquids-- this tensor is averaged out by rapid molecular tumbling, leading to a scalar interaction. Besides, for the same reason, the direct dipole-dipole coupling can be neglected.
\end{thebibliography}
\end{document}